\documentclass[twocolumn,showpacs,preprintnumbers]{revtex4}
\usepackage{graphicx}

\input{tcilatex}

\begin{document}

\preprint{123XXXX}

\textbf{Comment on \textquotedblleft Temporal scaling at Feigenbaum points
and non-extensive thermodynamics\textquotedblright }

In a recent letter [1], P. Grassberger addresses the very interesting issue
of the applicability of $q$-statistics to the renowned Feigenbaum attractor.
However several points are not in line with our current knowledge, nor are
the interpretations that he advances.

To begin with, contrary to the statement in [1] there is a simple relation
linking the constants in Eqs. (3) and (4). This is $\beta =d_{f}\log
_{2}\alpha $, where $d_{f}$ is the fractal dimension of the attractor. The
neglect of oscillations due to the multifractal nature of the attractor (his
Ref. [31]) in the rate $\overline{\Lambda }_{n}(1)$ leads to the above
equality. The derivation of Eq. (3) for $z>1$ is given in [2] so there is no
need to reproduce it in [1].

A more important issue lies behind the Author's comment that Eq. (3) holds
only for special values of time $n$. This is true as it is also true that
there are many other special values of $n$ that satisfy Eq. (3) exactly, all
with the same value of $q$. See Ref. 36 in [1] and [3]. All together these
sequences cover all $n$. Is there any key meaning behind this? As explained
[3], the dynamical organization within the attractor is difficult to resolve
from a simple time evolution: starting from an arbitrary position $x_{0}$ 
\textit{on} the attractor and recorded at every $n$. What is observed are
strong fluctuations that persist in time with a scrambled pattern structure.
Conversely, unsystematic averages over $x_{0}$ and/or $n$ would rub out the
details of the multiscale properties. However, if specific initial positions
with known location within the multifractal are chosen, and subsequent
positions are observed \textit{only} at pre-selected times, when the
trajectories visit another region of choice, a well-defined $q$-exponential
sensitivity appears, with $q$ and the associated Lyapunov spectrum $\lambda
(x_{0})$ determined by the attractor universal constants.

Another point in case is the suggestion in [1] of adopting Eq. (4) as focal
point for the natural generalization of the Lyapunov exponent. This has
already been considered in Refs. [28] and [29] in [1], although, yet again,
before taking a time average so that dynamical detail is preserved. A
straightforward calculation shows that%
\[
\lambda (x_{0}=1)\equiv \frac{1}{\ln n}\ln \left\vert \frac{dg}{dx_{0}}%
\right\vert =\frac{1}{n}\ln _{q}\left\vert \frac{dg}{dx_{0}}\right\vert
=\log _{2}\alpha , 
\]%
where $(1-q)^{-1}=\log _{2}\alpha $, and $\ln _{q}y$ is the $q$-logarithm,
the inverse of the $q$-exponential. So, the earlier definition for the
generalized $\lambda $ is equivalent to that given for the same quantity by
the $q$-statistics. The meaning of the index $q$ is given by the above
equalities. It is the degree of `$q$-deformation' of the ordinary logarithm
that makes $\lambda $ finite for large $n$. The physical origin of $q$ is
associated to the occurrence of dynamical phase transitions, of the
so-called Mori's $q$-phase type, as demonstrated in [3].

The identity derived in Ref. [36] of [1] between the rate of $q$-entropy
change and the generalized Lyapunov exponent is not the identity $%
(S_{n}-S_{0})n^{-1}=\Lambda _{n}n^{-1}$ in [1] (the zero identity for $%
n\rightarrow \infty $) but refers to $\lambda (x_{0})$ as above. Of course
it considers an instantaneous entropy rate $K(x_{0})$ (comparable in the
sense of [4] to the $q$-generalized KS entropy studied in Ref. [14] of [1]).
The identity $K(x_{0})=\lambda (x_{0})$ holds for $n\rightarrow \infty $ as
the interval length (around $x_{0}$) vanishes. It does fluctuate, but as
explained, we look for the detailed dependence on both $x_{0}$ and $n$. In
contrast to the chaotic case there is not one identity but many, and the
argument in [1] that averages are needed for applications of Pesin's
identity seems not to be useful for nonergodic and nonmixing trajectories.
Our results may not be insignificant as these can be reproduced [5]
combining the arguments in Ref. [14] of [1] regarding the $q$-KS entropy
with the results in [3]. One obtains the same equalities as for $\lambda
(x_{0}=1)$ above, with $\lambda (x_{0}=1)$ and $\left\vert
dg/dx_{0}\right\vert $ replaced by $K(x_{0}=1)$ and $\zeta _{n}(x_{0})$,
respectively, where $\zeta _{n}(x_{0})=Z_{n}^{1/1-q}$ and where $Z_{n}$ is
Mori's partition function [3]. On the contrary, the Renyi entropies $%
H_{n}^{q}$ in [1] from symbolic dynamics do not sense the universal constant 
$\alpha $ and/or the nonlinearity $z$.

On the subject of the `rich zoo' of $q$ values, there is a well-defined
family of these within the attractor, determined by the discontinuities of
the universal trajectory scaling function $\sigma $ [3]. There is a
corresponding family of Mori's $q$-phase transitions, each associated to
orbits that have common starting and finishing positions at specific
locations of the attractor. The special values for $q$ in the sensitivity
are equal to those of the variable q in the formalism of Mori \textit{et al}
at which the dynamical transitions take place [3]. Since the
discontinuities' amplitudes diminish rapidly, there is a hierarchical
structure in this family. The dominant discontinuity of $\sigma $ is
associated to the most crowded and most sparse regions in the attractor, and
this alone provides a reasonable description of the dynamics for which the
above expressions for $\lambda $ and $K$ belong. About generality, a very
similar picture has been recently obtained for another multifractal critical
attractor, that of the quasiperiodic route to chaos [6].

A strong reason for preferring a $q$-exponential to a power law does not
concern small arguments but the presence of a time scale factor (the
generalized $\lambda $) absent (or hidden) in the power law. This useful
quantity can be immediately `read' from the anomalous sensitivity just like
the ordinary $\lambda $ in chaotic dynamics. It is worth mentioning that the
(renormalization group) fixed-point map for intermittency, the other route
to chaos, is rigorously given by a $q$-exponential map. (See Ref. [19] in
[1]).

Alberto Robledo,*

Instituto de F\'{\i}sica, UNAM,

Apartado Postal 20-364,

M\'{e}xico 01000 D.F., Mexico

*Electronic address: robledo@fisica.unam.mx

{\small [1] P. Grassberger, Phys. Rev. Lett. 95, 140601 (2005).}

{\small [2] E. Mayoral, A. Robledo, Physica A 340, 219 (2004).}

{\small [3] E. Mayoral, A. Robledo, Phys Rev. E 72, 026209 (2005).}

{\small [4] V. Latora, M. Baranger, Phys Rev. Lett. 82, 520 (1999).}

{\small [5] H. Hern\'{a}ndez-Salda\~{n}a, A. Robledo, in preparation.}

{\small [6] H. Hern\'{a}ndez-Salda\~{n}a, A. Robledo, cond-mat/0507624.}

\end{document}